\documentclass[conference]{IEEEtran}
\IEEEoverridecommandlockouts
\usepackage{cite}
\usepackage{amsmath,amssymb,amsfonts}
\usepackage{algorithmic}
\usepackage{algorithm}
\usepackage{graphicx}
\usepackage[caption=false,font=footnotesize]{subfig}
\usepackage{textcomp}
\usepackage[dvipsnames,svgnames,table]{xcolor}
\usepackage{booktabs}
\usepackage{makecell}
\newcommand{\methodname}{MUTE}
\newcommand{\method}{\textbf{\texttt{\methodname}}}
\newcommand{\gray}[1]{\textcolor{gray}{#1}}
\newcommand{\venue}[1]{{\scriptsize\textcolor{gray!70!black}{[#1]}}}
\colorlet{bestcell}{red!13}
\colorlet{secondcell}{blue!10}
\newcommand{\best}[1]{\cellcolor{bestcell}\textbf{#1}}
\newcommand{\second}[1]{\cellcolor{secondcell}\textbf{#1}}
\DeclareRobustCommand{\swatch}[1]{{\setlength{\fboxsep}{0pt}\setlength{\fboxrule}{0.4pt}\raisebox{-0.2ex}{\fcolorbox{gray!80}{#1}{\rule{0pt}{1.8ex}\hspace{5ex}}}}}
\newcommand{\band}[1]{\gray{\textit{#1}}}
\definecolor{hdrA}{HTML}{8A1F3D}
\definecolor{hdrB}{HTML}{16479B}
\definecolor{hdrC}{HTML}{15683A}
\definecolor{hdrD}{HTML}{4D4D4D}
\newcommand{\hd}[2]{\cellcolor{#1}\textcolor{white}{\textbf{#2}}}
\definecolor{oursname}{HTML}{FBE5D0}
\newcommand{\std}[1]{{\scriptsize\textcolor{ForestGreen}{($\pm$#1)}}}
\usepackage{tcolorbox}
\usepackage{ntheorem}
\theoremseparator{:}

\usepackage{hyperref}
\usepackage{balance}

\def\BibTeX{{\rm B\kern-.05em{\sc i\kern-.025em b}\kern-.08em
    T\kern-.1667em\lower.7ex\hbox{E}\kern-.125emX}}
\begin{document}

\title{When Unlearning Fails: Reliable Data Deletion under Post-Training in Agent Networks}

\author{
\IEEEauthorblockN{Zihao Ding\textsuperscript{1}, Jun Huang\textsuperscript{1}, and Liang Dong\textsuperscript{2}}
\IEEEauthorblockA{\textsuperscript{1}Department of Electrical Engineering and Computer Science, South Dakota State University, Brookings, SD 57007, USA\\
\textsuperscript{2}Department of Electrical and Computer Engineering, Baylor University, Waco, TX 76798, USA\\
Email: zihao.ding@jacks.sdstate.edu, jun.huang@sdstate.edu, liang\_dong@baylor.edu}
}

\maketitle

\begin{abstract}
Embedded AI agents on edge devices can learn as a federation and keep improving after deployment. Each agent collects new trajectories with the current policy, and the edge server aggregates them in later rounds. In such self-improving agent networks, deleting data is harder than a one-time model repair. The data to be deleted may have already shaped later retained trajectories, so retraining or model-side unlearning leaves an influence echo that returns as the network keeps learning. We propose \underline{M}uting \underline{U}nlearned \underline{T}rajectories' \underline{E}choes (\method{}) for reliable deletion over edge networks. \method{} traces the echo from a lightweight ledger at the edge server, without moving raw trajectories off the devices. It erases the current residue with a forget-retain update, contains high-influence retained trajectories, and audits later behavior to schedule further erasure under the uplink budget and the response deadline.
On LIBERO, \method{} keeps the regenerated share of forgotten behavior at or below 0.401 for every request type, against at least 0.466 for five recent methods.
\end{abstract}


\section{Introduction}
\label{sec:introduction}

\IEEEPARstart{F}{ederated} learning (FL) is increasingly deployed as a long-running network service rather than a one-time training job~\cite{Wu2026TNSE,Wu2026COMST,Wu2026Network,Wu2026TON}. A central coordinator trains a model with many clients by exchanging model updates instead of collecting raw data, which is important in edge computing, Internet of Things (IoT), and cross-silo learning systems where data is private, distributed, and costly to move~\cite{Ding2026TAAS,Ding2025IPCCC,Wu2026IoTJ,Ding2026ICNC}. Modern edge agents further turn FL into a self-improving loop: they deploy the current model, interact with the environment, keep verified trajectories, and feed those data back into later rounds~\cite{Wu2026arXivRouting,Wu2026arXivREBASE,Wu2026arXivCrystalMem}. We call this setting a self-improving federated agent network.

Keeping raw data local does not settle the privacy obligation of such a network~\cite{Che2026arXiv}. A data owner may request that a trajectory, a client dataset, or a task source no longer influence the trained model. Regulations such as the General Data Protection Regulation (GDPR) give users a right to be forgotten~\cite{EU2016GDPR}. Federated unlearning (FU) addresses this requirement by removing the effect of requested data from a federated model~\cite{Romandini2025TNNLS,Ding2026Network}. Existing FU methods reduce deletion cost through rollback or model editing~\cite{Fraboni2024AISTATS,Pan2025AAAI,Zhong2025CVPR,Ding2026ICDCS,Pudasaini2026HPSR,Ding2026arXivEASE}. They mainly solve the request-time problem: once a request arrives, they update the current model to behave as if the target data had not been used.

\begin{figure}[t]
  \centering
  \includegraphics[width=0.95\columnwidth]{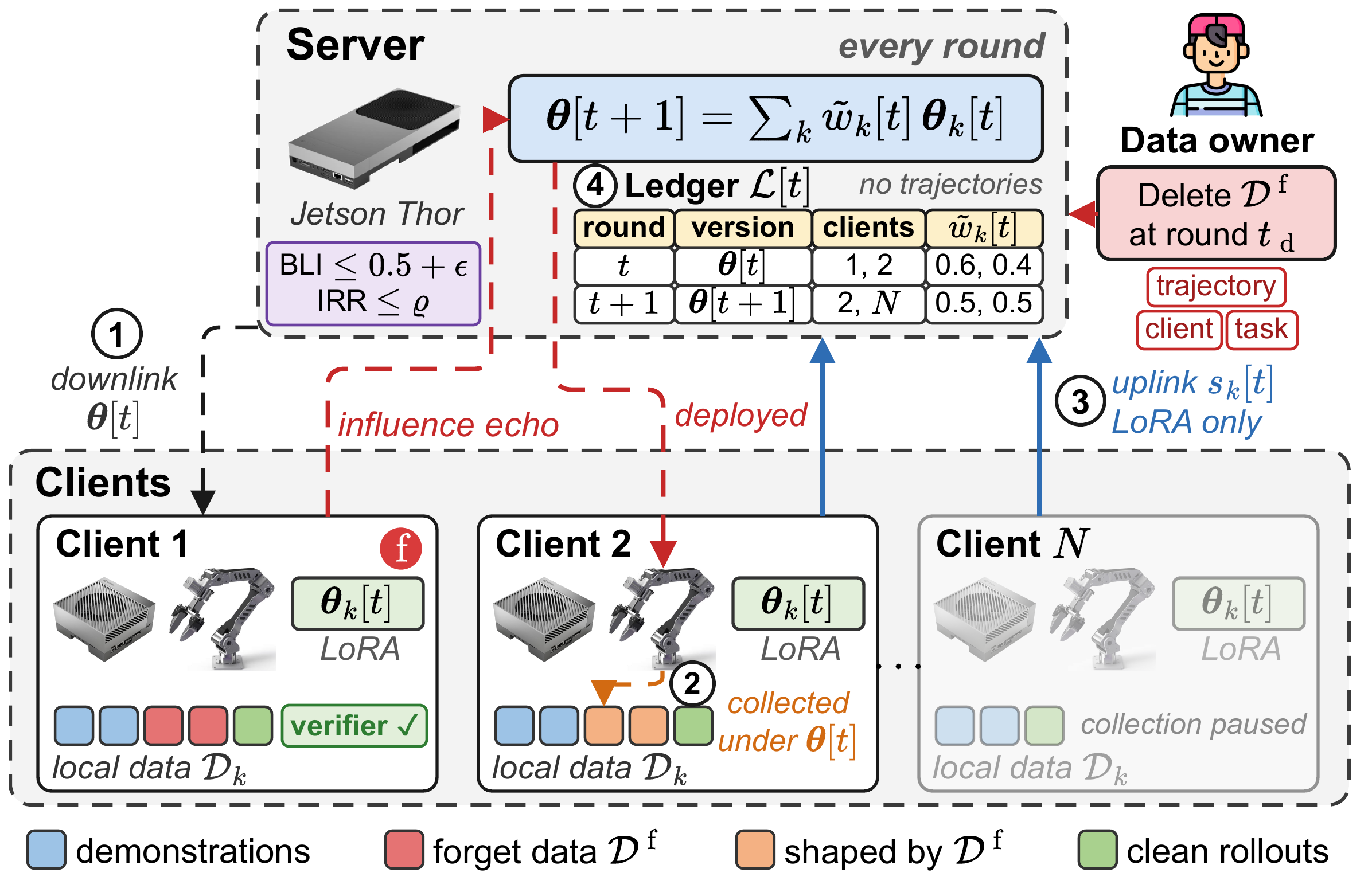}
  \caption{System model of a self-improving federated agent network.}
  \label{fig:system}
\end{figure}

This assumption breaks in a self-improving network. The deployed policy shapes what clients collect next, and those new trajectories are reused for later training~\cite{Shumailov2024Nature}. If the forget data shaped the policy before deletion, its influence may already have moved into retained data. Retraining on the observed retain set can therefore still train on data causally downstream of the forget set. After operation resumes, the forgotten behavior may return and its membership signal may become visible again~\cite{Carlini2022SP}. We call this phenomenon the influence echo of forgotten data. As shown in Fig.~\ref{fig:system}, reliable deletion must remove not only the direct data residue, but also the policy-driven echo that can regenerate later.

This leads to our central question: \emph{how can a federated network keep a deletion request valid while its agents continue collecting and training on new data?}

Prior work addresses only parts of this requirement. FU methods usually assume a fixed retain set and do not model policy-driven future data~\cite{Romandini2025TNNLS}. Continual and edge learning preserve useful knowledge across long-running tasks, but their goal is retention rather than sustained deletion~\cite{Wu2026TNSE,Wu2026ICDCS,Wu2026arXivLifecycle,Wu2026arXivPRISM}. Machine unlearning methods for large models can suppress target behavior with preference or gradient-based updates~\cite{Zhang2024COLM,Fan2025NeurIPS,Gao2025ICLR}, but model-side updates alone do not stop contaminated retained trajectories from reintroducing the behavior. Model-side deletion can also be undone by relearning attacks~\cite{Fan2025ICML}. Existing methods lack a unified mechanism that traces downstream influence, removes both model-side and data-side residues, and audits whether the influence reappears during continued operation.

To address these limitations, we propose \underline{\textbf{M}}uting \underline{\textbf{U}}nlearned \underline{\textbf{T}}rajectories' \underline{\textbf{E}}choes (\method{}), a reliable deletion method for self-improving federated agent networks. \method{} has three modules. It first replays a lightweight server ledger to estimate how the forget data's influence propagates through aggregation and policy-driven collection without moving raw trajectories off clients. It then performs influence-aware erasure, where a model-side update removes direct residue and a data-side step quarantines or down-weights high-influence retained trajectories. Finally, it audits behavioral leakage and influence regeneration, and schedules later erasure or containment actions under an uplink budget.

The main contributions of this paper are summarized as follows. We formulate reliable deletion for self-improving federated agent networks, where retain data may itself be shaped by forget data and deletion must remain valid after continued operation. We show that retraining can fail, regeneration scales with forget-shaped retained data, and the echo can be traced from network-side records. We develop \method{} to combine influence provenance, influence-aware erasure, and behavioral audit with scheduling. We validate \method{} on LIBERO with two VLA backbones and three deletion granularities, and further check its practicality on a physical Jetson-based edge testbed.
\section{System Model and Problem Formulation}
\label{sec:systemAndproblem}

\subsection{System Model}
\label{subsec:system-model}

We consider a centralized federated network with one server and $N$ heterogeneous clients. Each client $k$ holds a local dataset $\mathcal{D}_k$ that never leaves the client. The server and clients exchange model parameters instead of raw data, which preserves data locality and reduces network traffic. Let $\mathcal{D}=\bigcup_{k=1}^{N}\mathcal{D}_k$ denote the union of all local datasets. The federation trains a policy $\pi(\cdot;\boldsymbol{\theta})$ with parameters $\boldsymbol{\theta}$ by minimizing
\begin{equation}
    \min_{\boldsymbol{\theta}}\;\sum_{k=1}^{N}\frac{|\mathcal{D}_k|}{|\mathcal{D}|}\,\mathbb{E}_{\mathbf{x}\sim\mathcal{D}_k}\!\left[\ell(\mathbf{x};\boldsymbol{\theta})\right],
    \label{eq:fl-objective}
\end{equation}
where $\ell(\cdot)$ is the per-sample loss. At round $t$, the server sends the global parameter $\boldsymbol{\theta}[t]$ to selected clients. Each selected client $k$ updates it locally into $\boldsymbol{\theta}_k[t]$, and the server aggregates the uploaded parameters by
\begin{equation}
    \boldsymbol{\theta}[t+1]\;=\;\sum_{k=1}^{N}\tilde{w}_k[t]\,\boldsymbol{\theta}_k[t],
    \label{eq:fl-aggregation}
\end{equation}
where $\tilde{w}_k[t]=|\mathcal{D}_k|/|\mathcal{D}|$ is the normalized aggregation weight of client $k$.

Unlike a static federation, the network keeps improving while it operates. At round $t$, client $k$ deploys $\pi(\cdot;\boldsymbol{\theta}[t])$, acts in its environment, keeps the trajectories accepted by a verifier, and appends them to $\mathcal{D}_k$. The next local update in~\eqref{eq:fl-aggregation} then trains on the enlarged dataset. Thus, data collection depends on the deployed policy, while the deployed policy depends on all data collected so far.

We now introduce unlearning into this process. A data owner issues a deletion request that marks a forget set $\mathcal{D}^{\textrm{f}}\subseteq\mathcal{D}$, leaving the retain set $\mathcal{D}^{\textrm{r}}=\mathcal{D}\setminus\mathcal{D}^{\textrm{f}}$. We consider three granularities: a trajectory-level request removes selected sample trajectories, a client-level request removes the dataset of one client, and a task-level request removes all trajectories of one task across clients. The goal is to erase the effect of $\mathcal{D}^{\textrm{f}}$ while preserving the utility supported by $\mathcal{D}^{\textrm{r}}$.

In a static federation, this goal is well defined because the retain set is fixed: one removes $\mathcal{D}^{\textrm{f}}$ and retrains on $\mathcal{D}^{\textrm{r}}$. In our setting, this reference no longer holds. Since data collection is policy-driven, and the policy has been shaped by $\mathcal{D}^{\textrm{f}}$, the retain set $\mathcal{D}^{\textrm{r}}$ may also carry the influence of $\mathcal{D}^{\textrm{f}}$. Retraining on $\mathcal{D}^{\textrm{r}}$ may therefore fail to reach the counterfactual network $\boldsymbol{\theta}^{\star}$ that would have been produced had $\mathcal{D}^{\textrm{f}}$ never appeared, as Fig.~\ref{fig:echo} illustrates. Moreover, requests arrive while training continues, so erased influence may re-enter the model through later collection. We call this effect the regeneration of forgotten influence. Deletion must therefore be sustained over network operation, rather than enforced only once.

\subsection{Problem Formulation}
\label{subsec:problem-formulation}

We formulate deletion over the network above. Since the influence of $\mathcal{D}^{\textrm{f}}$ appears through behavior, not only through labeled test error, we measure current leakage by the behavioral leakage index $\mathrm{BLI}(\boldsymbol{\theta})$. It is the area under the ROC curve of a membership inference attack that distinguishes forget data from behavioral traces. We measure whether the leakage returns after continued learning by the influence regeneration rate $\mathrm{IRR}$. We write $U(\boldsymbol{\theta})$ for task utility and $\tau$ for the response time of a request. At each round, a client either keeps computation local or uploads an adapter update. Let $s_k[t]$ denote the uplink volume of client $k$ at round $t$, which equals one adapter size when the client transmits and zero otherwise.

The counterfactual network $\boldsymbol{\theta}^{\star}$ is the ideal clean-deletion reference, but it cannot be reproduced in a live network because the past cannot be re-collected without $\mathcal{D}^{\textrm{f}}$. A deletion method must instead drive leakage to the chance level and keep it there while the network continues to learn. At each round, the method chooses client actions $\mathbf{a}[t]$ for tracing, erasure, containment, and audit. We minimize the communication cost of deleting $\mathcal{D}^{\textrm{f}}$ while preserving utility, meeting the response deadline, and preventing regeneration:
\begin{equation}
\begin{split}
    (\mathrm{P}):\;\;\min_{\{\mathbf{a}[t]\}}\;\;
    & \sum_{t=1}^{T}\sum_{k=1}^{N} s_k[t] \\
    \mathrm{s.t.}\;\;
    & (\mathrm{C}_1):\; \mathrm{BLI}(\boldsymbol{\theta}[t]) \le 0.5+\epsilon, \\
    & (\mathrm{C}_2):\; U(\boldsymbol{\theta}[t]) \ge (1-\delta)\,U_0, \\
    & (\mathrm{C}_3):\; \tau \le \tau_{\max}, \\
    & (\mathrm{C}_4):\; \mathrm{IRR} \le \varrho,
\end{split}
\label{eq:mute-problem}
\end{equation}
where $\epsilon$ is the chance-level margin, $U_0$ is the pre-request utility, $\delta$ is the tolerated utility drop, $\tau_{\max}$ is the response deadline, and $\varrho$ is the tolerated regeneration after $T$ rounds. Constraint $(\mathrm{C}_1)$ limits current behavioral leakage, and $(\mathrm{C}_2)$ preserves task utility. Constraint $(\mathrm{C}_3)$ enforces timely response, while $(\mathrm{C}_4)$ requires deletion to remain valid after continued learning.
\section{Sustaining Network Privacy with \method{}}
\label{sec:method}

\begin{figure*}[t]
  \centering
  \begin{minipage}[t]{0.48\textwidth}
    \centering
    \includegraphics[width=\linewidth]{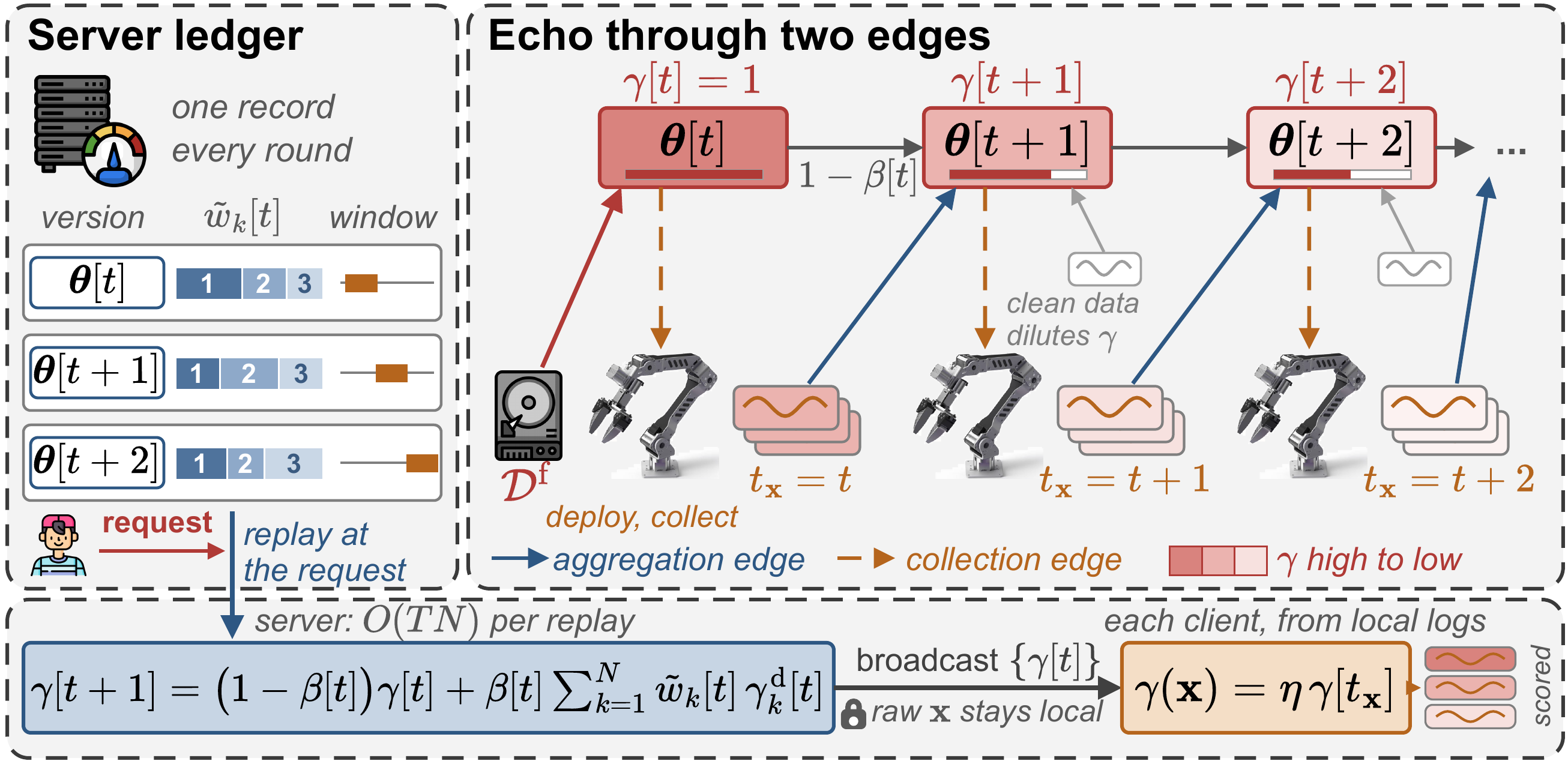}
    \caption{Influence provenance by ledger replay.}
    \label{fig:provenance}
  \end{minipage}
  \hfill
  \begin{minipage}[t]{0.48\textwidth}
    \centering
    \includegraphics[width=\linewidth]{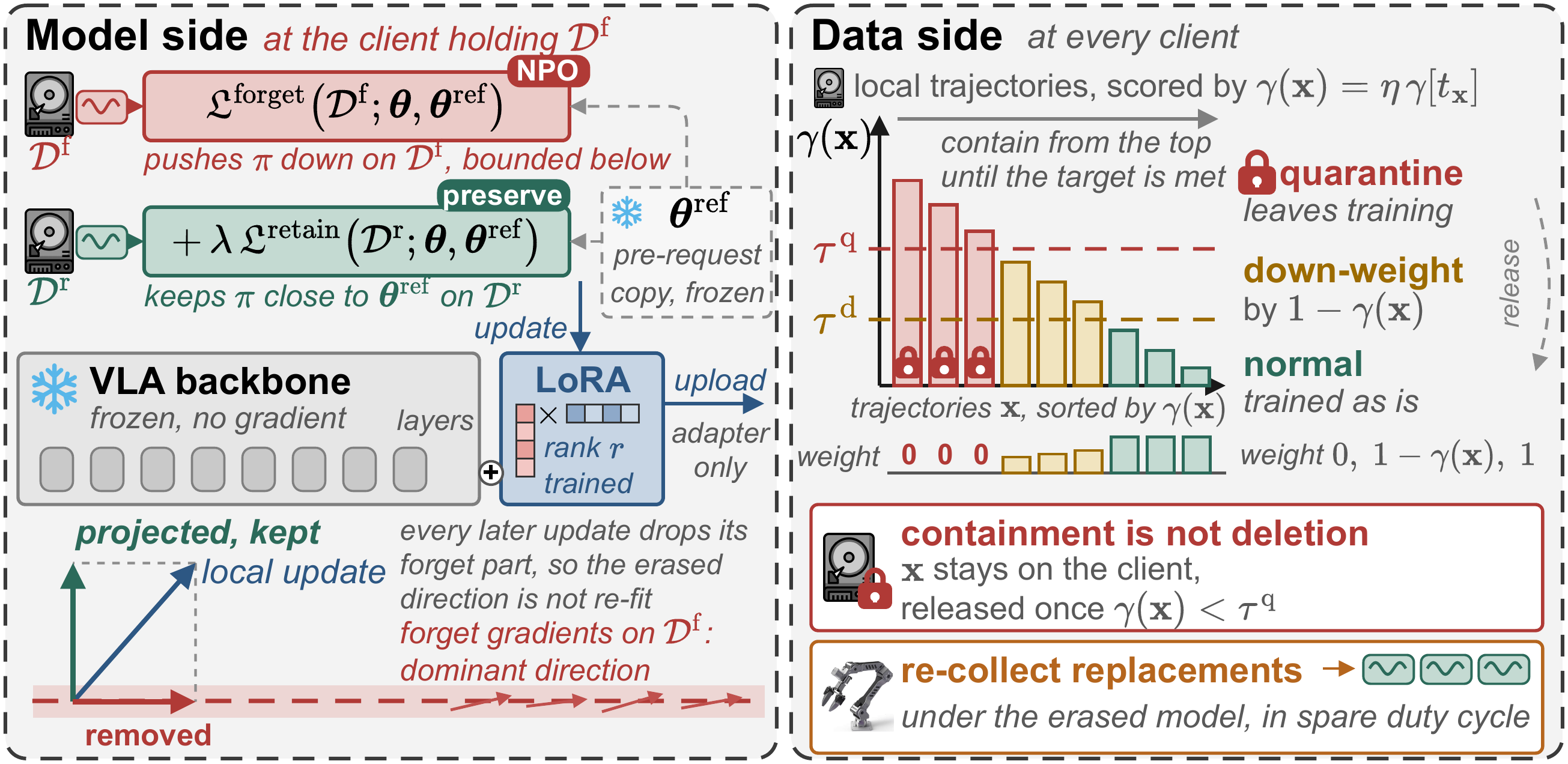}
    \caption{Influence-aware erasure on both sides.}
    \label{fig:erasure}
  \end{minipage}
\end{figure*}

To solve the problem in~\eqref{eq:mute-problem}, we design \underline{\textbf{M}}uting \underline{\textbf{U}}nlearned \underline{\textbf{T}}rajectories' \underline{\textbf{E}}choes (\method{}), an unlearning method for self-improving networks. \method{} follows three steps. It traces how the forget data's influence spreads through aggregation and data collection, erases the influence from both the model and high-risk retained trajectories, and sustains deletion through behavioral audit and scheduling.

\begin{figure*}[t]
  \centering
  \begin{minipage}[t]{0.48\textwidth}
    \centering
    \includegraphics[width=\linewidth]{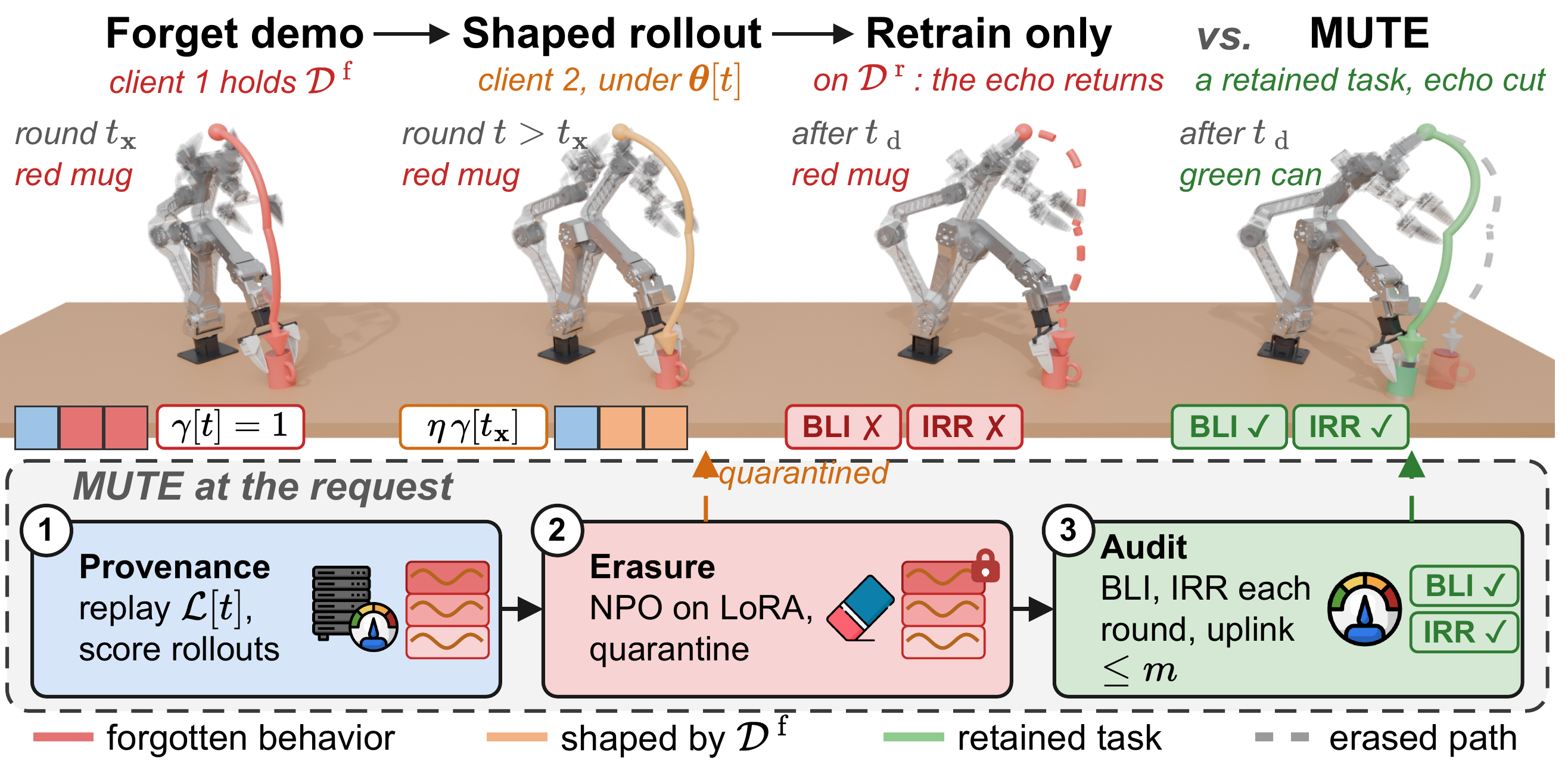}
    \caption{How \method{} cuts the influence echo.}
    \label{fig:echo}
  \end{minipage}
  \hfill
  \begin{minipage}[t]{0.48\textwidth}
    \centering
    \includegraphics[width=\linewidth]{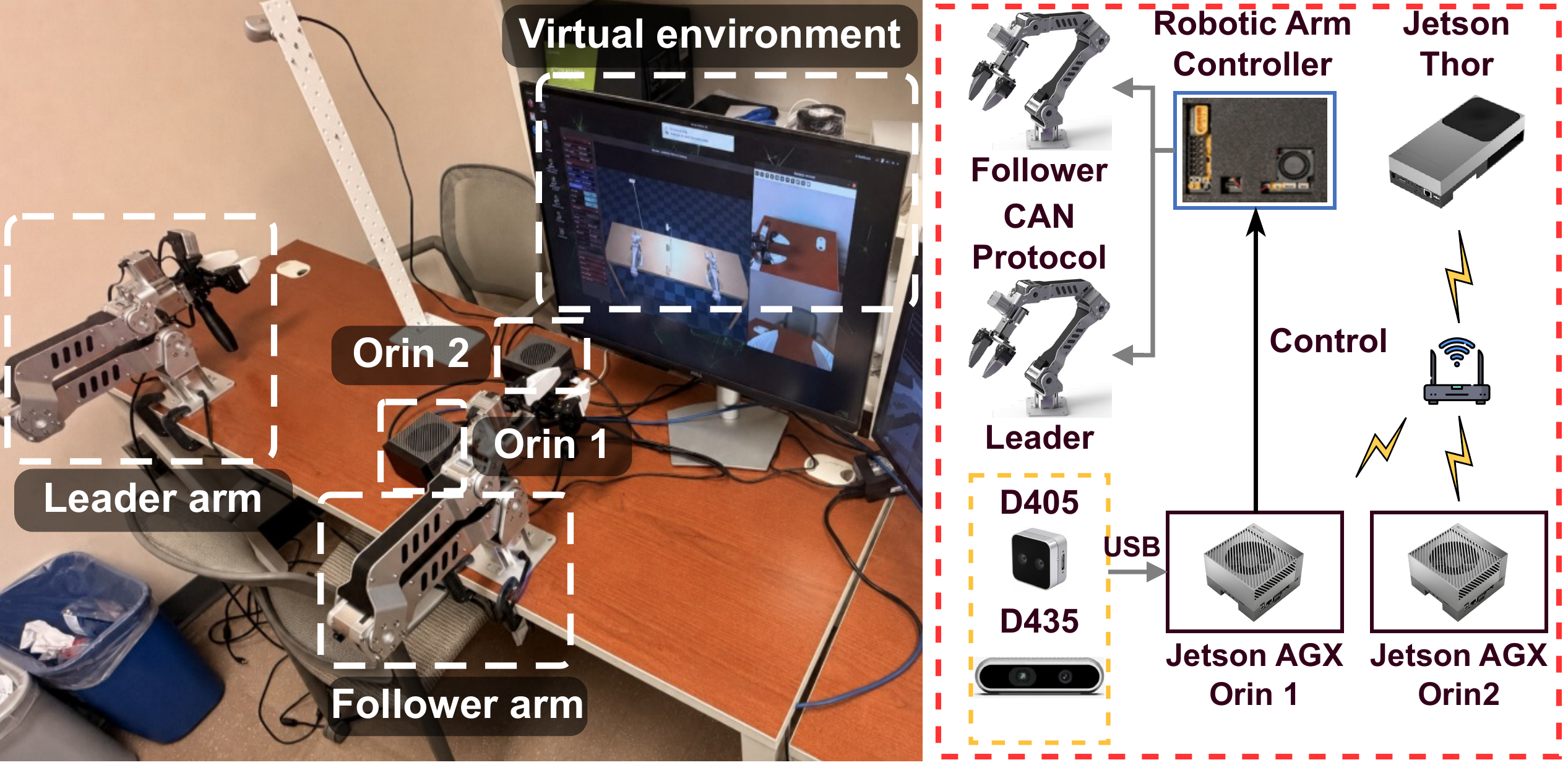}
    \caption{Physical testbed and its topology.}
    \label{fig:testbed}
  \end{minipage}
\end{figure*}

\subsection{Influence Provenance}
\label{subsec:provenance}

Retained trajectories collected under a shaped policy may inherit the influence of $\mathcal{D}^{\textrm{f}}$. This module converts that influence into a network-side score. Classical data attribution estimates influence from per-sample gradients or Hessians~\cite{Koh2017ICML}, which the server cannot compute without the raw trajectories. During normal operation, the server records each round's deployed model version, participating clients, aggregation weights, and deployment windows. We denote this lightweight ledger by $\mathcal{L}[t]$. When an unlearning request arrives, the server replays the ledger to estimate where the influence has propagated.

The influence spreads through two edges. The aggregation edge carries influence from local training data to the global model and then to other clients. We set $\gamma[t]=1$ at the round where $\mathcal{D}^{\textrm{f}}$ first enters training, and update the influence of later global versions by
\begin{equation}
    \gamma[t{+}1] =
    \bigl(1-\beta[t]\bigr)\gamma[t]
    + \beta[t]\sum_{k=1}^{N}\tilde{w}_k[t]\gamma_k^{\textrm{d}}[t],
    \label{eq:version-influence}
\end{equation}
where $\gamma_k^{\textrm{d}}[t]\in[0,1]$ is the mean influence of client $k$'s current training data, $\tilde{w}_k[t]$ is its normalized aggregation weight, and $\beta[t]$ is the relative size of the round update.

The collection edge carries influence from a deployed model to the trajectories collected under it. If trajectory $\mathbf{x}$ is collected under version $\boldsymbol{\theta}[t_{\mathbf{x}}]$, its influence score is
\begin{equation}
    \gamma(\mathbf{x}) = \eta\,\gamma[t_{\mathbf{x}}],
    \label{eq:traj-influence}
\end{equation}
where $\eta\in[0,1]$ captures how strongly the deployed policy shapes data collection. Since the ledger follows the round order, one replay costs $O(TN)$. The server only broadcasts the scalar sequence $\{\gamma[t]\}$, and each client scores its own trajectories from local collection logs using~\eqref{eq:traj-influence}. Raw trajectories and per-trajectory metadata remain local.

The two edges feed each other through the data term of Eq.~\eqref{eq:version-influence}. Each client returns one scalar, the weighted mean of its trajectory scores,
\begin{equation}
    \gamma_k^{\textrm{d}}[t] =
    \frac{\sum_{\mathbf{x}\in\mathcal{D}_k}\omega(\mathbf{x})\,\gamma(\mathbf{x})}
         {\sum_{\mathbf{x}\in\mathcal{D}_k}\omega(\mathbf{x})},
    \label{eq:data-influence}
\end{equation}
where the training weight $\omega(\mathbf{x})$ is one unless lowered by containment in Section~\ref{subsec:erasure}. Substituting Eq.~\eqref{eq:data-influence} into Eq.~\eqref{eq:version-influence} closes the loop. A version shaped by $\mathcal{D}^{\textrm{f}}$ passes its influence to the data it collects, which pass it to the next version.

The version influence $\gamma[t]$ decays on its own, since a round trained only on clean data adds nothing to the second term of Eq.~\eqref{eq:version-influence}. The factor $\eta$ is an upper bound rather than a fitted constant. Over-scoring costs utility that later collection can win back, while under-scoring misses a carrier.

For scheduling, we summarize the current influence as a network state. Let $\mathbf{x}[t]$ collect the model-side and data-side influence of all clients. Aggregation and collection move influence across this state, while containment and erasure reduce it. We write
\begin{equation}
    \mathbf{x}[t{+}1]=\mathbf{M}(\mathbf{a}[t])\,\mathbf{x}[t],
    \label{eq:influence-dynamics}
\end{equation}
where $\mathbf{M}(\mathbf{a}[t])$ is the transition matrix under actions $\mathbf{a}[t]$.

\subsection{Influence-Aware Erasure}
\label{subsec:erasure}

This module removes the current residue from the model and blocks future residue from retained trajectories.

For model-side erasure, the client holding $\mathcal{D}^{\textrm{f}}$ updates its adapter with a forget-retain objective:
\begin{equation}
    \mathfrak{L}
    =
    \mathfrak{L}^{\textrm{forget}}
    \bigl(\mathcal{D}^{\textrm{f}};\boldsymbol{\theta},\boldsymbol{\theta}^{\textrm{ref}}\bigr)
    +
    \lambda\,
    \mathfrak{L}^{\textrm{retain}}
    \bigl(\mathcal{D}^{\textrm{r}};\boldsymbol{\theta},\boldsymbol{\theta}^{\textrm{ref}}\bigr),
    \label{eq:npo}
\end{equation}
where $\mathfrak{L}^{\textrm{forget}}$ suppresses behavior on forget trajectories, $\mathfrak{L}^{\textrm{retain}}$ preserves behavior on retained data relative to the pre-request reference model $\boldsymbol{\theta}^{\textrm{ref}}$, and $\lambda$ balances unlearning and utility. In implementation, $\mathfrak{L}^{\textrm{forget}}$ follows negative preference optimization (NPO)~\cite{Zhang2024COLM}. The update touches only the low-rank adapter, so each upload remains small. To avoid later recovery, subsequent local updates are projected away from the dominant forget-gradient subspace~\cite{Saha2021ICLR}. Let the columns of $\mathbf{V}$ be the top singular vectors of the adapter gradients of $\mathfrak{L}^{\textrm{forget}}$ at the end of erasure. Every later local update $\Delta\boldsymbol{\theta}_k[t]=\boldsymbol{\theta}_k[t]-\boldsymbol{\theta}[t]$ is replaced by
\begin{equation}
    \Delta\boldsymbol{\theta}_k[t] \leftarrow
    \bigl(\mathbf{I}-\mathbf{V}\mathbf{V}^{\top}\bigr)\Delta\boldsymbol{\theta}_k[t],
    \label{eq:projection}
\end{equation}
so the erased direction receives no further gradient. $\mathbf{V}$ is computed once per request and broadcast on the downlink.

We use NPO because gradient ascent has no lower bound and breaks the retained behavior first. Projection is needed because retained trajectories are correlated with $\mathcal{D}^{\textrm{f}}$, so an update without Eq.~\eqref{eq:projection} can re-fit the erased direction.

For data-side containment, each client grades its trajectories by $\gamma(\mathbf{x})$. Trajectories with $\gamma(\mathbf{x})\ge\tau^{\textrm{q}}$ are quarantined and removed from later training. Trajectories with $\tau^{\textrm{d}}\le\gamma(\mathbf{x})<\tau^{\textrm{q}}$ are down-weighted by $1-\gamma(\mathbf{x})$, while those with $\gamma(\mathbf{x})<\tau^{\textrm{d}}$ are used normally. In Eq.~\eqref{eq:data-influence}, $\omega(\mathbf{x})$ is therefore zero, $1-\gamma(\mathbf{x})$, and one in the three bands. Containment is not deletion. A quarantined trajectory stays on its client and only leaves training, so it is released once its score drops below $\tau^{\textrm{q}}$.

\subsection{Behavioral Audit and Deletion Scheduling}
\label{subsec:schedule}

A one-time erasure is not enough because the network continues to learn after the request. Behavior planted in a federated model is known to persist over many later rounds~\cite{Bagdasaryan2020AISTATS}. This module audits the residue and schedules later erasure, containment, audit, and normal training under resource limits.

We use the two audit signals of Section~\ref{subsec:metrics}. BLI checks the current model and passes below $0.5+\epsilon$. IRR checks whether high-influence retained data revive the forgotten behavior. Audited clients return scores, not trajectories.

At each round, the server assigns each client normal training, unlearning, containment, audit, or paused collection under the uplink budget, response deadline, and utility floor, favoring clients with large model-side or data-side influence.

The feasibility of sustained unlearning follows from~\eqref{eq:influence-dynamics}. The influence decays, and $(\mathrm{C}_4)$ becomes satisfiable, when the closed-loop transition matrix $\mathbf{M}^{\infty}$ under the long-run schedule satisfies $\rho(\mathbf{M}^{\infty})<1$. The server therefore selects the lowest-communication schedule that meets the deadline and utility constraints while keeping this stability condition.

The search is small in practice. The server ranks clients by their entries of $\mathbf{x}[t]$ and assigns erasure and containment from the top within the budget $m$. It then evaluates $\rho(\mathbf{M}^{\infty})$ for this schedule. If $\rho(\mathbf{M}^{\infty})\ge 1$, the next-ranked client moves from normal training to containment, and the check repeats. Each check is an eigenvalue computation on a $2N\times 2N$ matrix, negligible at the server. Once both signals pass, all clients resume normal training under periodic audit.

\begin{table*}[!t]
\centering
\setlength{\tabcolsep}{4pt}
\setlength{\aboverulesep}{0pt}
\setlength{\belowrulesep}{0pt}
\setlength{\extrarowheight}{0pt}
\renewcommand{\arraystretch}{0.88}
\caption{Deletion results per backbone and request type, as the mean over five seeds with the standard deviation in parentheses. BLI is ideal at 0.5, and Comm is in MB. A \swatch{bestcell} cell marks the best value and a \swatch{secondcell} cell the second best within each backbone, excluding Oracle.}
\label{tab:main}
\resizebox{\linewidth}{!}{\begin{tabular}{l|cccc|cccc|cccc}
\Xhline{1.2pt}
 & \multicolumn{4}{c|}{\hd{hdrA}{Trajectory}} & \multicolumn{4}{c|}{\hd{hdrB}{Client}} & \multicolumn{4}{c}{\hd{hdrC}{Task}}\\
\textbf{Method} & SR$\uparrow$ & BLI & IRR$\downarrow$ & Comm$\downarrow$ & SR$\uparrow$ & BLI & IRR$\downarrow$ & Comm$\downarrow$ & SR$\uparrow$ & BLI & IRR$\downarrow$ & Comm$\downarrow$\\
\Xhline{1.2pt}
\multicolumn{13}{l}{\band{(a) MiniVLA}}\\
\Xhline{0.6pt}
Retrain & \best{0.452\std{0.016}} & \best{0.357\std{0.053}} & 0.931\std{0.033} & 220.0 & \best{0.562\std{0.030}} & \second{0.479\std{0.028}} & 0.802\std{0.015} & 176.0 & 0.450\std{0.021} & 0.295\std{0.037} & 1.000 & 220.0\\
FedOSD \venue{AAAI'25} & \second{0.449\std{0.019}} & 0.814\std{0.058} & 0.867\std{0.012} & 52.8 & 0.437\std{0.041} & 0.625\std{0.022} & 0.730\std{0.053} & 44.0 & 0.454\std{0.017} & 0.666\std{0.021} & \second{0.967\std{0.018}} & 79.2\\
FUSED \venue{CVPR'25} & 0.421\std{0.033} & 0.791\std{0.061} & 0.931\std{0.017} & \second{22.0} & 0.472\std{0.022} & 0.703\std{0.049} & 0.853\std{0.011} & \second{17.6} & 0.450\std{0.023} & 0.675\std{0.021} & 1.000 & \best{22.0}\\
SimNPO \venue{NeurIPS'25} & 0.422\std{0.041} & 0.845\std{0.066} & \second{0.844\std{0.035}} & 44.0 & \second{0.501\std{0.025}} & 0.708\std{0.048} & 0.809\std{0.021} & 44.0 & 0.447\std{0.018} & 0.690\std{0.033} & 1.000 & 44.0\\
SAM-NPO \venue{ICML'25} & 0.411\std{0.019} & 0.784\std{0.039} & 0.972\std{0.013} & 44.0 & 0.470\std{0.039} & 0.754\std{0.047} & 0.809\std{0.049} & 44.0 & 0.514\std{0.038} & \second{0.571\std{0.069}} & 1.000 & 44.0\\
O3 \venue{ICLR'25} & 0.440\std{0.025} & 0.663\std{0.026} & \second{0.844\std{0.070}} & \best{8.8} & 0.433\std{0.025} & 0.588\std{0.051} & \second{0.702\std{0.054}} & \best{8.8} & \second{0.536\std{0.030}} & \best{0.442\std{0.057}} & 1.000 & \second{35.2}\\
\Xhline{0.8pt}
\cellcolor{oursname}\textit{\textbf{\methodname{}}} & 0.418\std{0.028} & \second{0.659\std{0.023}} & \best{0.317\std{0.026}} & 88.0 & 0.405\std{0.019} & \best{0.517\std{0.045}} & \best{0.118\std{0.055}} & 176.0 & \best{0.547\std{0.026}} & 0.606\std{0.039} & \best{0.000} & 44.0\\
\Xhline{0.8pt}
\textit{Oracle} & 0.413\std{0.020} & 0.322\std{0.024} & 0.623\std{0.067} & - & 0.493\std{0.034} & 0.426\std{0.054} & 0.160\std{0.019} & - & 0.446\std{0.018} & 0.315\std{0.066} & 0.000 & -\\
\Xhline{0.8pt}
\multicolumn{13}{l}{\band{(b) $\pi_0$}}\\
\Xhline{0.6pt}
Retrain & \best{0.759\std{0.052}} & \best{0.509\std{0.097}} & 0.645\std{0.066} & 973.2 & \best{0.718\std{0.036}} & \second{0.487\std{0.025}} & 0.619\std{0.055} & 778.6 & \best{0.713\std{0.040}} & \best{0.481\std{0.059}} & 0.810\std{0.042} & 973.2\\
FedOSD \venue{AAAI'25} & 0.628\std{0.059} & 0.640\std{0.080} & 0.849\std{0.063} & 233.6 & 0.618\std{0.058} & 0.606\std{0.034} & 0.577\std{0.068} & 194.6 & 0.614\std{0.050} & 0.655\std{0.035} & 0.854\std{0.041} & 350.4\\
FUSED \venue{CVPR'25} & 0.636\std{0.046} & 0.770\std{0.054} & 0.810\std{0.089} & \second{97.3} & 0.597\std{0.055} & 0.672\std{0.053} & 0.726\std{0.162} & \second{77.9} & 0.555\std{0.054} & \second{0.602\std{0.079}} & 0.899\std{0.111} & \best{97.3}\\
SimNPO \venue{NeurIPS'25} & \second{0.644\std{0.041}} & 0.766\std{0.038} & 0.526\std{0.070} & 194.6 & \second{0.624\std{0.037}} & 0.704\std{0.038} & \second{0.550\std{0.119}} & 194.6 & 0.680\std{0.091} & 0.658\std{0.055} & 0.601\std{0.139} & 194.6\\
SAM-NPO \venue{ICML'25} & 0.615\std{0.029} & 0.742\std{0.055} & 0.881\std{0.035} & 194.6 & 0.623\std{0.064} & 0.710\std{0.060} & 0.575\std{0.055} & 194.6 & \best{0.713\std{0.027}} & 0.688\std{0.041} & 0.761\std{0.067} & 194.6\\
O3 \venue{ICLR'25} & 0.604\std{0.065} & \second{0.596\std{0.044}} & \second{0.466\std{0.068}} & \best{38.9} & 0.550\std{0.057} & \best{0.507\std{0.064}} & 0.751\std{0.053} & \best{38.9} & 0.590\std{0.067} & 0.618\std{0.042} & \second{0.587\std{0.074}} & \second{155.7}\\
\Xhline{0.8pt}
\cellcolor{oursname}\textit{\textbf{\methodname{}}} & 0.612\std{0.019} & 0.632\std{0.060} & \best{0.401\std{0.046}} & 280.3\std{111.5} & 0.573\std{0.035} & 0.582\std{0.110} & \best{0.150\std{0.031}} & 599.5\std{112.2} & 0.666\std{0.042} & 0.642\std{0.036} & \best{0.068\std{0.046}} & 280.3\std{57.7}\\
\Xhline{0.8pt}
\textit{Oracle} & 0.666\std{0.049} & 0.459\std{0.079} & 0.443\std{0.054} & - & 0.669\std{0.040} & 0.476\std{0.052} & 0.177\std{0.049} & - & 0.768\std{0.032} & 0.538\std{0.055} & 0.105\std{0.058} & -\\
\Xhline{1.2pt}
\end{tabular}}
\end{table*}
\begin{figure*}[!t]
\centering
\subfloat[Regeneration over time.]{\includegraphics[width=0.24\textwidth]{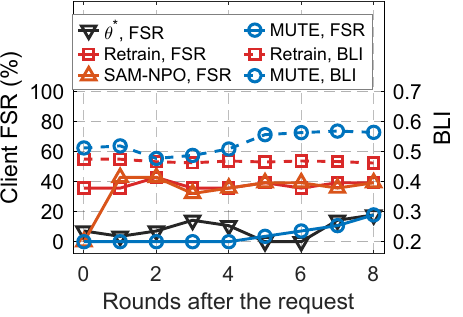}\label{fig:longrun}}
\hfill
\subfloat[Source of the echo.]{\includegraphics[width=0.24\textwidth]{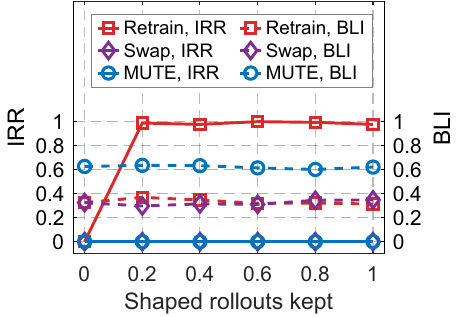}\label{fig:echosrc}}
\hfill
\subfloat[Score accuracy.]{\includegraphics[width=0.24\textwidth]{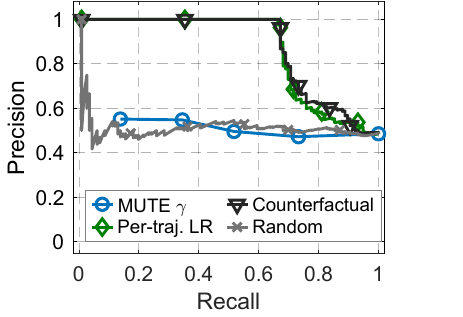}\label{fig:score}}
\hfill
\subfloat[Shaping factor.]{\includegraphics[width=0.24\textwidth]{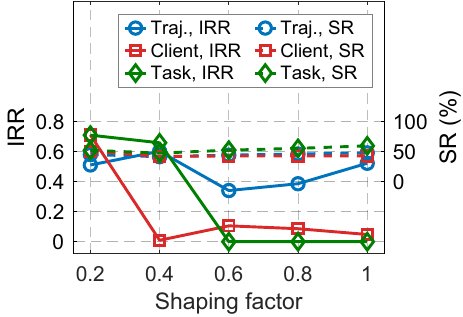}\label{fig:eta}}
\caption{Influence echo and influence score.}
\label{fig:results}
\end{figure*}

\section{Performance Evaluation}
\label{sec:exp}

\subsection{Experimental Setup}
\label{subsec:setup}

We fine-tune two vision-language-action policies with low-rank adapters of rank 8 on frozen backbones. MiniVLA has one billion parameters and emits discrete action tokens, and $\pi_0$ has three billion parameters and produces continuous actions by flow matching. The ten tasks of LIBERO-Goal are spread over $N=5$ clients by a Dirichlet distribution with concentration 0.4. Each client seeds its data with the benchmark demonstrations and keeps the rollouts that pass the task-success check. A request arrives after five rounds and names one trajectory set, one client, or one task, and we follow the network for $K=8$ further rounds. \method{} uses a shaping factor $\eta=0.6$, thresholds $\tau^{\textrm{q}}=0.15$ and $\tau^{\textrm{d}}=0.05$, and an uplink budget of four clients per round.

We compare \method{} with the five most recent methods on the same rounds, adapters, and uplink accounting. FedOSD~\cite{Pan2025AAAI} unlearns by orthogonal steepest descent and then recovers, and FUSED~\cite{Zhong2025CVPR} retrains sparse adapters on the retained data. SimNPO~\cite{Fan2025NeurIPS} and SAM-NPO~\cite{Fan2025ICML} run a forget-retain update at the client that holds the forget data, the latter with sharpness-aware steps against relearning. O3~\cite{Gao2025ICLR} adds one orthogonal adapter per request behind a gate. Retrain trains again from the start on the retained data, and Oracle never sees the forget data or anything collected under a policy it shaped and serves as the reference. On the testbed in Fig.~\ref{fig:testbed}, a Jetson Thor server and two Jetson AGX Orin clients run the same loop on real Panthera-HT arms over a LAN, where we measure the uplink bytes and the response time to a request. Further clients hosted on the Thor bring the network to $N=2$, 4, and 8.

\subsection{Performance Metrics}
\label{subsec:metrics}

We evaluate deletion from four aspects: task utility, current leakage, future regeneration, and communication overhead. The counterfactual network $\boldsymbol{\theta}^{\star}$ is the ideal reference. Since it cannot be reproduced in a live network, we compare BLI with the chance level $0.5$ and IRR with zero.

\subsubsection{Success Rate}
Success rate (SR) measures task utility. It is the accepted fraction of rollouts, $\mathrm{SR}(\boldsymbol{\theta}[t])=\mathbb{E}_{\mathbf{x}\sim\pi(\cdot;\boldsymbol{\theta}[t])}[v(\mathbf{x})]$, where the verifier $v(\mathbf{x})=1$ accepts a successful trajectory. Rollouts run on the retained tasks.

\begin{figure*}[!t]
\centering
\subfloat[Ablation.]{\includegraphics[width=0.24\textwidth]{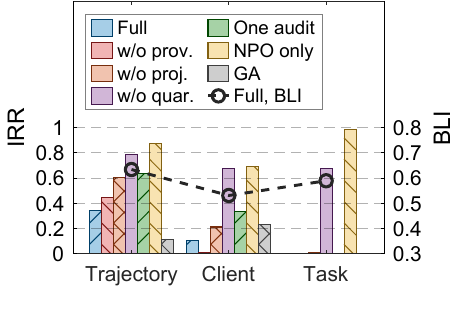}\label{fig:ablation}}
\hfill
\subfloat[Testbed.]{\includegraphics[width=0.24\textwidth]{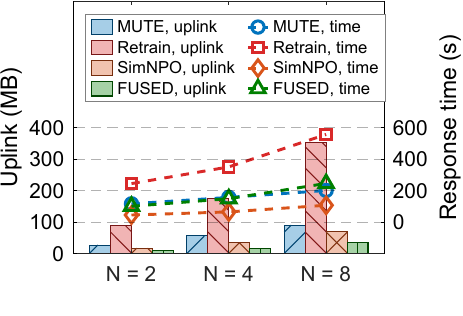}\label{fig:tbres}}
\hfill
\subfloat[What is forgotten.]{\includegraphics[width=0.24\textwidth]{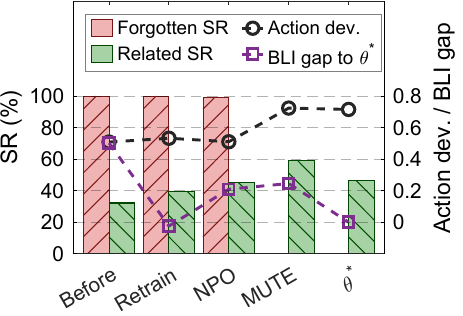}\label{fig:forget}}
\hfill
\subfloat[Consent given again.]{\includegraphics[width=0.24\textwidth]{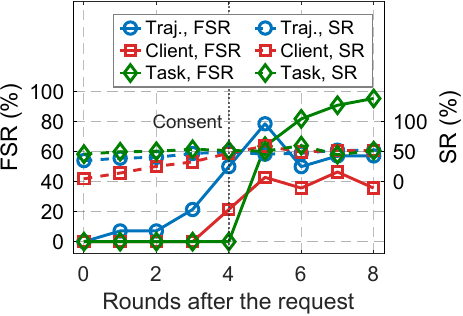}\label{fig:consent}}
\caption{Ablation, testbed, and scope of deletion.}
\label{fig:ablation-testbed}
\end{figure*}

\begin{figure}[t]
\centering
\subfloat[Quarantine threshold.]{\includegraphics[width=0.485\linewidth]{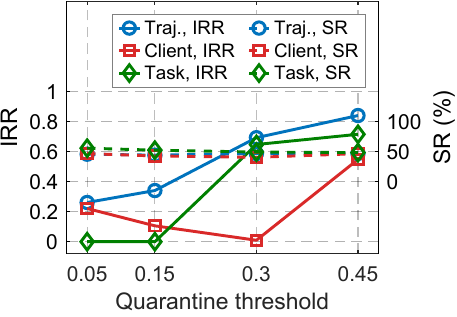}\label{fig:tauq}}
\hfill
\subfloat[Uplink budget.]{\includegraphics[width=0.485\linewidth]{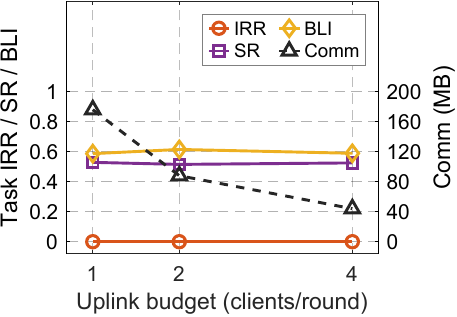}\label{fig:budget}}
\caption{Quarantine threshold and uplink budget.}
\label{fig:sensitivity}
\end{figure}

\subsubsection{Behavioral Leakage Index}
Behavioral leakage index (BLI) measures how much the forget data still leaks through behavior. Following the membership inference attack~\cite{Carlini2022SP}, we use the per-sample loss $\ell(\mathbf{x};\boldsymbol{\theta})$ as the membership score on canary queries in the task stream. Let $\mathcal{D}^{\textrm{o}}$ be a held-out non-member set. BLI is defined as
\begin{equation}
    \mathrm{BLI}(\boldsymbol{\theta}[t])=\frac{\bigl|\{(\mathbf{x},\mathbf{x}')\in\mathcal{D}^{\textrm{f}}\times\mathcal{D}^{\textrm{o}}:\ell(\mathbf{x};\boldsymbol{\theta}[t])<\ell(\mathbf{x}';\boldsymbol{\theta}[t])\}\bigr|}{|\mathcal{D}^{\textrm{f}}|\,|\mathcal{D}^{\textrm{o}}|}.
    \label{eq:bli}
\end{equation}
This is the area under the ROC curve of the attack. The ideal value is the chance level $0.5$, and a higher BLI indicates stronger leakage. Constraint $(\mathrm{C}_1)$ bounds it.

\subsubsection{Influence Regeneration Rate}
Influence regeneration rate (IRR) measures whether deleted behavior returns as learning continues. Let $\mathrm{FSR}(\boldsymbol{\theta})$ denote the forget success rate, the rate at which the deleted behavior is reproduced under model $\boldsymbol{\theta}$. The pre-request reference model $\boldsymbol{\theta}^{\textrm{ref}}$ still has the behavior, and $\boldsymbol{\theta}^{\textrm{unl}}[t]$ is the network $t$ rounds after the request. IRR is the mean recovered fraction over the $K$ rounds,
\begin{equation}
    \mathrm{IRR}=\frac{1}{K}\sum_{t=1}^{K}\frac{\mathrm{FSR}(\boldsymbol{\theta}^{\textrm{unl}}[t])}{\mathrm{FSR}(\boldsymbol{\theta}^{\textrm{ref}})}.
    \label{eq:irr}
\end{equation}
IRR is $0$ when the behavior stays erased and $1$ when it fully returns. Constraint $(\mathrm{C}_4)$ bounds it.

\subsubsection{Communication Cost}
Communication cost (Comm) measures the extra uplink traffic introduced by deletion. Normal training already uploads adapter updates, so we count only the additional uploads for tracing, erasure, containment, and audit after the request, summed over clients and rounds.

\subsection{Experimental Results}
\label{subsec:results}

In Table~\ref{tab:main}, \method{} attains the lowest IRR at every granularity on both backbones. On MiniVLA, its IRR is 0.317, 0.118, and 0.000 for trajectory, client, and task requests, against at least 0.844, 0.702, and 0.967 for the baselines. On $\pi_0$, it lowers IRR to 0.401, 0.150, and 0.068. \method{} trails Retrain in SR by up to 0.157, and its BLI stays between 0.517 and 0.659. Audits raise its client-level Comm to 176.0~MB.

In Fig.~\ref{fig:longrun}, Retrain and SAM-NPO return to a client-level FSR of 32\% to 43\% within one round. \method{} stays at zero for five rounds and ends at 18\%, equal to $\boldsymbol{\theta}^{\star}$. In Fig.~\ref{fig:echosrc}, keeping 20\% of the shaped rollouts lifts the task-level IRR of Retrain to 0.989, while $\boldsymbol{\theta}^{\star}$ replays from the same states keep it at zero. The echo thus comes from the shaped data. In Fig.~\ref{fig:score}, the version-level score ranks affected rollouts at an AUC of 0.51, near random, while a per-trajectory likelihood ratio reaches 0.80. Containment works because $\tau^{\textrm{q}}$ quarantines almost every rollout collected after $\mathcal{D}^{\textrm{f}}$ entered training. Fig.~\ref{fig:eta} shows that $\eta=0.6$ is the smallest value that clears the task-level echo.

Without quarantine, IRR rises to 0.78, 0.67, and 0.68 in Fig.~\ref{fig:ablation}, and NPO alone gives 0.88, 0.69, and 0.99. Uniform scores and gradient ascent each beat the full method at one granularity. On the testbed in Fig.~\ref{fig:tbres}, \method{} uploads a quarter to a third of the Retrain traffic and answers in 116 to 201~s, against 245 to 557~s. In Fig.~\ref{fig:forget}, \method{} drives the forgotten task to zero success with an action deviation of 0.724, close to 0.715 for $\boldsymbol{\theta}^{\star}$. After consent is given again, FSR returns within four rounds in Fig.~\ref{fig:consent}. In Fig.~\ref{fig:sensitivity}, raising $\tau^{\textrm{q}}$ to 0.45 lifts trajectory-level IRR to 0.84, and a larger uplink budget cuts Comm from 176 to 44~MB.

\section{Conclusion}
\label{sec:conclusion}

We studied reliable data deletion in self-improving federated agent networks, where a deletion request must remain effective while deployed policies continue collecting new training data. One-time unlearning is insufficient because the forget data leaves an influence echo in later retained trajectories. \method{} addresses this problem by tracing downstream influence from server-side records, erasing model-side residue, containing high-influence retained trajectories, and auditing later regeneration under an uplink budget.
On LIBERO, \method{} attains the lowest regeneration of forgotten behavior at all three request granularities on both backbones. On a Jetson testbed, it answers a request two to three times faster than retraining.



\begin{thebibliography}{10}
\providecommand{\url}[1]{#1}
\csname url@samestyle\endcsname
\providecommand{\newblock}{\relax}
\providecommand{\bibinfo}[2]{#2}
\providecommand{\BIBentrySTDinterwordspacing}{\spaceskip=0pt\relax}
\providecommand{\BIBentryALTinterwordstretchfactor}{4}
\providecommand{\BIBentryALTinterwordspacing}{\spaceskip=\fontdimen2\font plus
\BIBentryALTinterwordstretchfactor\fontdimen3\font minus
  \fontdimen4\font\relax}
\providecommand{\BIBforeignlanguage}[2]{{%
\expandafter\ifx\csname l@#1\endcsname\relax
\typeout{** WARNING: IEEEtran.bst: No hyphenation pattern has been}%
\typeout{** loaded for the language `#1'. Using the pattern for}%
\typeout{** the default language instead.}%
\else
\language=\csname l@#1\endcsname
\fi
#2}}
\providecommand{\BIBdecl}{\relax}
\BIBdecl

\bibitem{Wu2026TNSE}
B.~Wu, Z.~Ding, and J.~Huang, ``{A Review of Continual Learning in Edge AI},''
  \emph{IEEE Transactions on Network Science and Engineering}, vol.~13, pp.
  6571--6588, 2026.

\bibitem{Wu2026COMST}
B.~Wu, J.~Huang, and S.~Yu, ````{X} of information'' continuum: A survey on
  {AI}-driven multi-dimensional metrics for next-generation networked
  systems,'' \emph{{IEEE} Communications Surveys \& Tutorials}, vol.~28, pp.
  5307--5344, 2026.

\bibitem{Wu2026Network}
B.~Wu, J.~Huang, and Q.~Duan, ``Real-time intelligent healthcare enabled by
  federated digital twins with {AoI} optimization,'' \emph{{IEEE} Network},
  vol.~40, no.~2, pp. 184--191, 2026.

\bibitem{Wu2026TON}
B.~Wu, J.~Huang, Q.~Duan, L.~Dong, and Z.~Cai, ``Enhancing vehicular platooning
  with wireless federated learning: A resource-aware control framework,''
  \emph{{IEEE} Transactions on Networking}, vol.~34, pp. 1479--1494, 2026.

\bibitem{Ding2026TAAS}
Z.~Ding, J.~Huang, Y.~Zhao, and Z.~Cai, ``Combating knowledge diversity and
  catastrophic forgetting in uav-assisted collaborative vehicular learning: A
  game-theoretic approach,'' \emph{ACM Trans. Auton. Adapt. Syst.}, Jun. 2026,
  just Accepted.

\bibitem{Ding2025IPCCC}
Z.~Ding, J.~Huang, Q.~Duan, C.~Zhang, Y.~Zhao, and S.~Gu, ``A dual-level
  game-theoretic approach for collaborative learning in {UAV}-assisted
  heterogeneous vehicle networks,'' in \emph{2025 {IEEE} International
  Performance, Computing, and Communications Conference ({IPCCC})}, 2025, pp.
  1--8.

\bibitem{Wu2026IoTJ}
B.~Wu, Z.~Ding, and J.~Huang, ``{RELIEF}: Turning missing modalities into
  training acceleration for federated learning on heterogeneous {IoT} edge,''
  \emph{{IEEE} Internet of Things Journal}, 2026, early access.

\bibitem{Ding2026ICNC}
Z.~Ding, J.~Huang, and J.~Qi, ``Learning to defend: A multi-agent reinforcement
  learning framework for {Stackelberg} security game in mobile edge
  computing,'' in \emph{2026 International Conference on Computing, Networking
  and Communications ({ICNC})}, 2026, pp. 769--774.

\bibitem{Wu2026arXivRouting}
B.~Wu, Z.~Ding, and J.~Huang, ``Not all experience belongs in the weights:
  Component routing for self-improving {GUI} agents,'' \emph{arXiv preprint
  arXiv:2610.01787}, 2026.

\bibitem{Wu2026arXivREBASE}
B.~Wu, J.~Huang, and Y.~Zhao, ``{REBASE}: Device-cloud experience coherence for
  {GUI} agents across app updates,'' \emph{arXiv preprint arXiv:2609.32130},
  2026.

\bibitem{Wu2026arXivCrystalMem}
B.~Wu and J.~Huang, ``{CrystalMem}: Elastic memory for self-evolving {LLM}
  agents via knowledge crystallization,'' \emph{arXiv preprint
  arXiv:2608.00303}, 2026.

\bibitem{Che2026arXiv}
X.~Che, D.~Xu, and X.~Wu, ``{StraightDP}: Geometry-aware differential privacy
  for rectified-flow transformers,'' \emph{arXiv preprint arXiv:2607.29100},
  2026.

\bibitem{EU2016GDPR}
{European Parliament and Council of the European Union}, ``Regulation ({EU})
  2016/679 of the european parliament and of the council of 27 april 2016 on
  the protection of natural persons with regard to the processing of personal
  data and on the free movement of such data, and repealing directive
  95/46/{EC} (general data protection regulation),'' Official Journal of the
  European Union, {OJ} L 119, pp. 1--88, 2016.

\bibitem{Romandini2025TNNLS}
N.~Romandini, A.~Mora, C.~Mazzocca, R.~Montanari, and P.~Bellavista,
  ``Federated unlearning: {A} survey on methods, design guidelines, and
  evaluation metrics,'' \emph{{IEEE} Trans. Neural Networks Learn. Syst.},
  vol.~36, no.~7, pp. 11\,697--11\,717, 2025.

\bibitem{Ding2026Network}
Z.~Ding and J.~Huang, ``Toward trustworthy federated unlearning for mobile
  autonomous systems,'' \emph{IEEE Network}, pp. 1--9, 2026.

\bibitem{Fraboni2024AISTATS}
Y.~Fraboni, M.~V. Waerebeke, K.~Scaman, R.~Vidal, L.~Kameni, and M.~Lorenzi,
  ``{SIFU:} sequential informed federated unlearning for efficient and provable
  client unlearning in federated optimization,'' in \emph{International
  Conference on Artificial Intelligence and Statistics, 2-4 May 2024, Palau de
  Congressos, Valencia, Spain}, ser. Proceedings of Machine Learning Research,
  S.~Dasgupta, S.~Mandt, and Y.~Li, Eds., vol. 238.\hskip 1em plus 0.5em minus
  0.4em\relax {PMLR}, 2024, pp. 3457--3465.

\bibitem{Pan2025AAAI}
Z.~Pan, Z.~Wang, C.~Li, K.~Zheng, B.~Wang, X.~Tang, and J.~Zhao, ``Federated
  unlearning with gradient descent and conflict mitigation,'' in
  \emph{Proceedings of the {AAAI} Conference on Artificial Intelligence},
  vol.~39, no.~19, 2025, pp. 19\,804--19\,812.

\bibitem{Zhong2025CVPR}
Z.~Zhong, W.~Bao, J.~Wang, S.~Zhang, J.~Zhou, L.~Lyu, and W.~Y.~B. Lim,
  ``Unlearning through knowledge overwriting: Reversible federated unlearning
  via selective sparse adapter,'' in \emph{Proceedings of the {IEEE/CVF}
  Conference on Computer Vision and Pattern Recognition ({CVPR})}, 2025, pp.
  30\,661--30\,670.

\bibitem{Ding2026ICDCS}
Z.~Ding, B.~Wu, and J.~Huang, ``{Scale}: Sensitivity-aware federated unlearning
  with information freshness optimization for mobile edge computing,'' in
  \emph{2026 {IEEE} 46th International Conference on Distributed Computing
  Systems ({ICDCS})}, 2026, pp. 1476--1486.

\bibitem{Pudasaini2026HPSR}
U.~Pudasaini, Z.~Ding, and J.~Huang, ``Securing smart agriculture with
  communication-efficient federated unlearning,'' in \emph{2026 {IEEE} 27th
  International Conference on High Performance Switching and Routing ({HPSR})},
  2026, pp. 1--7.

\bibitem{Ding2026arXivEASE}
Z.~Ding, B.~Wu, and J.~Huang, ``{EASE}: Federated multimodal unlearning via
  entanglement-aware anchor closure,'' \emph{arXiv preprint arXiv:2605.00733},
  2026.

\bibitem{Shumailov2024Nature}
I.~Shumailov, Z.~Shumaylov, Y.~Zhao, N.~Papernot, R.~J. Anderson, and Y.~Gal,
  ``{AI} models collapse when trained on recursively generated data,''
  \emph{Nat.}, vol. 631, no. 8022, pp. 755--759, 2024.

\bibitem{Carlini2022SP}
N.~Carlini, S.~Chien, M.~Nasr, S.~Song, A.~Terzis, and F.~Tram{\`{e}}r,
  ``Membership inference attacks from first principles,'' in \emph{43rd {IEEE}
  Symposium on Security and Privacy, {SP} 2022, San Francisco, CA, USA, May
  22-26, 2022}.\hskip 1em plus 0.5em minus 0.4em\relax {IEEE}, 2022, pp.
  1897--1914.

\bibitem{Wu2026ICDCS}
B.~Wu, J.~Huang, and Y.~Zhao, ``From {Alpha} to {Omega}: Lifecycle-aware
  forgetting defense in federated continual learning for planetary
  exploration,'' in \emph{2026 {IEEE} 46th International Conference on
  Distributed Computing Systems ({ICDCS})}, 2026, pp. 1--11.

\bibitem{Wu2026arXivLifecycle}
B.~Wu and J.~Huang, ``Lifecycle-aware federated continual learning in mobile
  autonomous systems,'' \emph{arXiv preprint arXiv:2604.20745}, 2026.

\bibitem{Wu2026arXivPRISM}
B.~Wu, Z.~Ding, and J.~Huang, ``{PRISM}: Exposing and resolving spurious
  isolation in federated multimodal continual learning,'' \emph{arXiv preprint
  arXiv:2605.01061}, 2026.

\bibitem{Zhang2024COLM}
R.~Zhang, L.~Lin, Y.~Bai, and S.~Mei, ``Negative preference optimization: From
  catastrophic collapse to effective unlearning,'' in \emph{First Conference on
  Language Modeling}, 2024.

\bibitem{Fan2025NeurIPS}
C.~Fan, J.~Liu, L.~Lin, J.~Jia, R.~Zhang, S.~Mei, and S.~Liu, ``Simplicity
  prevails: Rethinking negative preference optimization for {LLM} unlearning,''
  in \emph{Advances in Neural Information Processing Systems ({NeurIPS})},
  2025.

\bibitem{Gao2025ICLR}
C.~Gao, L.~Wang, K.~Ding, C.~Weng, X.~Wang, and Q.~Zhu, ``On large language
  model continual unlearning,'' in \emph{The Thirteenth International
  Conference on Learning Representations, {ICLR} 2025, Singapore, April 24-28,
  2025}.\hskip 1em plus 0.5em minus 0.4em\relax OpenReview.net, 2025.

\bibitem{Fan2025ICML}
C.~Fan, J.~Jia, Y.~Zhang, A.~Ramakrishna, M.~Hong, and S.~Liu, ``Towards {LLM}
  unlearning resilient to relearning attacks: A sharpness-aware minimization
  perspective and beyond,'' in \emph{Proceedings of the 42nd International
  Conference on Machine Learning ({ICML})}, ser. Proceedings of Machine
  Learning Research, vol. 267, 2025, pp. 15\,762--15\,778.

\bibitem{Koh2017ICML}
P.~W. Koh and P.~Liang, ``Understanding black-box predictions via influence
  functions,'' in \emph{Proceedings of the 34th International Conference on
  Machine Learning, {ICML} 2017, Sydney, NSW, Australia, 6-11 August 2017},
  ser. Proceedings of Machine Learning Research, D.~Precup and Y.~W. Teh, Eds.,
  vol.~70.\hskip 1em plus 0.5em minus 0.4em\relax {PMLR}, 2017, pp. 1885--1894.

\bibitem{Saha2021ICLR}
G.~Saha, I.~Garg, and K.~Roy, ``Gradient projection memory for continual
  learning,'' in \emph{9th International Conference on Learning
  Representations, {ICLR} 2021, Virtual Event, Austria, May 3-7, 2021}.\hskip
  1em plus 0.5em minus 0.4em\relax OpenReview.net, 2021.

\bibitem{Bagdasaryan2020AISTATS}
E.~Bagdasaryan, A.~Veit, Y.~Hua, D.~Estrin, and V.~Shmatikov, ``How to backdoor
  federated learning,'' in \emph{The 23rd International Conference on
  Artificial Intelligence and Statistics, {AISTATS} 2020, 26-28 August 2020,
  Online [Palermo, Sicily, Italy]}, ser. Proceedings of Machine Learning
  Research, S.~Chiappa and R.~Calandra, Eds., vol. 108.\hskip 1em plus 0.5em
  minus 0.4em\relax {PMLR}, 2020, pp. 2938--2948.

\end{thebibliography}
\end{document}